\documentclass[10pt,aps,prl,showpacs,twocolumn]{revtex4-1}
\usepackage{amsmath,amssymb,graphicx}
\usepackage{mathptmx}
\usepackage{color}

\usepackage{xcolor}

\usepackage{textcomp}
\newcommand{\mkm}{\textmu m} %microns

\begin{document}

\title{Hydrodynamic 2D turbulence and spatial beam condensation in multimode optical fibers}

\author{E.\,V.~Podivilov$^{1,2}$}
\author{D.\,S.~Kharenko$^{1,2}$}
\author{V.\,A.~Gonta$^1$}
\author{K.~Krupa$^3$}
\author{O.\,S.~Sidelnikov$^{1,4}$}
\author{S.~Turitsyn$^{1,5}$}
\author{M.\,P.~Fedoruk$^{1,4}$}
\author{S.\,A.~Babin$^{1,2}$}
\author{S.~Wabnitz$^{1,6}$}

\affiliation{$^1$ Novosibirsk State University, Novosibirsk 630090, Russia}
\affiliation{$^2$ Institute of Automation and Electrometry SB RAS, 1 ac. Koptyug ave., Novosibirsk 630090, Russia}
\affiliation{$^3$ Dipartimento di Ingegneria dell'Informazione, Universit\`a di Brescia, Via Branze 38, 25123 Brescia, Italy}
\affiliation{$^4$ Institute of Computational Technologies SB RAS, Novosibirsk 630090, Russia}
\affiliation{$^5$ Aston Institute of Photonic Technologies, Aston University, Birmingham, B4 7ET, UK}
\affiliation{$^6$ Dipartimento di Ingegneria dell'Informazione, Elettronica e Telecomunicazioni, Sapienza Universit\`a di Roma, Via Eudossiana 18, 00184 Roma, Italy}

\date{\today}

\begin{abstract}
We show that Kerr beam self-cleaning results from parametric mode mixing instabilities, that generate a number of nonlinearly interacting modes with randomised phases -- optical wave turbulence, followed by a direct and inverse cascade towards high mode numbers and condensation into the fundamental mode, respectively. This optical self-organization effect is analogue to wave condensation that is well-known in hydrodynamic 2D turbulence.
\end{abstract}

\pacs{}
\keywords{}

\maketitle


Spatiotemporal light beam dynamics in multimode optical fibers (MMFs) has emerged in recent years as fertile research domain in nonlinear optics  \cite{Wright2015R31,Wright2015R29}. 
Intriguing spatiotemporal wave propagation phenomena such as multimode optical solitons \cite{Hasegawa} and parametric instabilities leading to ultra-wideband sideband series \cite{Longhi2003R27}, although predicted quite a long time ago, have only been experimentally observed in MMFs in the last few years \cite{Renninger2012R31,Wright2015R31, Wright2015R29,KrupaPRLGPI}. 
It is well known that linear wave propagation in MMFs is affected by random mode coupling, which leads to highly irregular speckled intensity patterns at the fiber output, even when the fiber is excited with a high quality, diffraction limited input beam. 
%The inability of multimode optical fibers to maintain the spatial quality of optical beams has largely prevented their technological applications so far.
Control and management of the spatial properties of optical beams in the multimode optical fibers is critically important to their various technological applications such as e.g. space-division multiplexing \cite{Richardson} aiming to increase the capacity of optical networks

Recent experiments \cite{KrupaPRLGPI,Krupanatphotonics,LiuKerr,WrightNP2016} have surprisingly discovered that the intensity dependent contribution to the refractive index, or Kerr effect, has the capacity to counteract such random mode coupling in a graded index (GRIN) MMF, leading to the formation of a highly robust beam. The self-cleaned beam at the fiber output has a size that is close to the fundamental mode of the MMF, and sits on a background of higher-order modes (HOMs).  Typically, spatial self-cleaning is observed in several meters of GRIN MMF at threshold power levels of the order of few kWs, orders of magnitude lower than the value for catastrophic self-focusing. Moreover, self-cleaning is most easily observed in a quasi-continuous wave (CW) propagation regime (i.e., by using sub-nanosecond pulses), so that dispersive effects can be neglected \cite{Krupanatphotonics,KrupaPRLGPI,WrightNP2016,Galmiche:16,KrupaLuot:16}. In such regime, nonlinear mode coupling is the sole mechanism responsible for the self-cleaning of the transverse spatial beam profile at the fiber output \cite{PhysRevA.97.043836}.

So far, although different qualitative explanations have been tentatively provided \cite{PRA11b, Krupanatphotonics,WrightNP2016,Laegsgaard:18}, the physical mechanism leading to Kerr beam cleaning remains largely debated. Full scale numerical simulations involving the full 3D+1 nonlinear Schroedinger equation (or Gross-Pitaievskii equation) are able to qualitatively reproduce the experimental observations, provided that the random mode coupling process is suitably modeled. 

In this Letter, we experimentally demonstrate that spatial beam self-cleaning in MMFs is fully analogous to hydrodynamic 2D turbulence, where a large-scale condensate (e.g., a hurricane, the Jupiter's red spot, etc.) results from parametric instabilities leading to a spatial mode redistribution process. As first outlined by Kraichnan  \cite{kraichnan,kandm}, the key ingredient of hydrodynamic 2D (as opposed to Kolmogorov's 3D) turbulence is the simultaneous presence of a direct energy cascade towards high wave numbers, accompanied by the inverse cascade towards low wave numbers and, ultimately, the condensate. Crucially, both direct and inverse energy cascades concur in conserving the average mode number throughout the whole process of development of the 2D turbulence. This is counter-intuitive, as the occurrence of condensation (or beam-self-cleaning) would lead one to expect a significant reduction of the average mode number. We present, we believe for the first time, a direct experimental demonstration of the conservation of the average mode number in the process of spatial beam cleaning. 

Kerr beam cleaning is a two-step physical process. In a first step, and above a certain input power threshold, four-wave mixing (FWM) among the spatial modes leads to a net power transfer from low-order modes (LOMs) towards the fundamental mode of the fiber. We demonstrate that this process is accompanied, for symmetry reasons, by a power transfer towards HOMs, while leaving the average mode number unchanged. 
We are able to analytically predict such power threshold by approximating the mode coupling process by a truncated three-mode expansion. This leads to an exactly integrable model \cite{Cappellini}: the power threshold corresponds to a spatial parametric instability associated with a bifurcation of the nonlinear system's eigenmodes. Instability results when the strength of FWM overcomes the diffractive or propagation constant mismatch between the modes. 

In a realistic situation, power transfer occurs between a large number of nonlinearly interacting modes with randomised phases, leading to optical wave turbulence, 
A cascaded four-wave mixing process occurs, which redistributes the input beam power away from LOMs, towards HOMs (direct cascade) and the fundamental mode (inverse cascade). The second step of self-cleaning is the nonlinear nonreciprocity-induced irreversible decoupling of the fundamental mode from HOMs\cite{Krupanatphotonics}. We confirm such a picture by full numerical simulations including all relevant modes and random mode coupling.

The equation for the field amplitude $E = A(z, \vec r) \exp(i\omega t - ik z)$ in a GRIN fiber (with refractive index $n(z, \vec r) = n_0\sqrt{1- \Delta^2 r^2} +n_2I +\delta n(z, \vec r) $) reads as
\begin{align}
%\label{E1}
&-2ik\frac{d A}{d z} + \frac{d^2 A}{d\vec r^2} - k^2 \Delta^2 r^2 A = -2k^2(n_2/n_0)|A|^2 A -\\ \nonumber
&\frac{d^2 A}{dz^2} - k^2(2\delta n(z, \vec r)/n_0)A 
%+ 2ik \vec q(z, \vec r) \frac{d A}{d\vec r}
,
\end{align}
where $\Delta $ is the mode spacing, $k=2\pi n_0/\lambda$ is the wave vector, and $\lambda$ is the wavelength. Terms in the right-hand side denote nonlinearity, 
angular dispersion, and random refractive index fluctuations.
$|A|^2=I$ is the intensity of the light beam, and all terms in RHS supposed to be small.
Let us introduce the following dimensionless and normalized variables
\begin{gather*}
	\vec \rho = \vec r/r_0,\quad r_0 =1/\sqrt{k\Delta}, \zeta = z \Delta, \\
	\delta (z, \vec r) = (2k\delta n(z, \vec r)/\Delta n_0), A=\Psi \frac{\sqrt{P}}{r_0}, \\
	p = \int |A|^2 d^2\vec{r}/P_{sf} = (P/P_{sf})\int |\Psi|^2 d\vec{\rho}^2 = (P/P_{sf}), 
\end{gather*}
where $P_{sf} = n_0/2n_2 k^2 \simeq 1 MW$ is the power threshold for catastrophic self-focusing. 
The equation for the normalized envelope $\Psi$ reads as
\begin{eqnarray}
	\label{E1}
	2i\frac{d\Psi}{d\zeta} = \frac{d^2 \Psi}{d\vec{\rho}^2 } -\rho^2\Psi + D\frac{d^2 \Psi}{d\zeta^2} 
	+ p |\Psi|^2\Psi + \delta(z, \vec r)\Psi
\end{eqnarray}
where $D=\Delta/k =\lambda/2L_B \ll 1 $ is the angular dispersion coefficient.
The normalized field envelope reads as
\begin{equation}
	\label{envelope}
	\Psi(\zeta,\vec \rho) = \sum\limits_{m,p=0}^\infty  B_{p,m}(\zeta) U_{p,m}(\vec \rho) e^{i (n+1) \zeta} 
	\end{equation}
where
\begin{equation}
	U_{p,m}(\vec \rho) = N_{p,m}\rho^{|m|} L^{|m|}_p(\rho^2) e^{-\rho^2/2} e^{i m \phi} \nonumber
\end{equation}

Here $L^{|m|}_p$ are Laguerre polynomials, $U_{p,m}(\vec \rho)$~-- normalized on unity transversal eigen mode, $N_{p,m}$~-- normalized coefficient and $n = 2 p + |m|$.
By inserting Eq.~\eqref{envelope} into Eq.~\eqref{E1}, one obtains equations for the complex amplitudes $B_{p,m}$.
After keeping only resonant four-wave mixing (FWM) terms, which do not rapidly oscillate with a beat length 
$z= L_B =\pi/\Delta$, one obtains
\begin{align} \label{Enumeric}
	2i\frac{dB_{p,m(\zeta)}}{d\zeta} \equiv D(n + 1)^2 B_{p,m} \\ \nonumber
	+ p\sum\limits_{m+m_1 = m_2 + m_3} \sum\limits_{p+p_1 = p_2 + p_3} f_{p,p_1,p_2,p_3}^{m,m_1,m_2,m_3} B_{p_1,m_1}^{*} B_{p_2,m_2} B_{p_3,m_3} \\  \nonumber
	+ \sum\limits_{m_1,p_1} C_{m,m_1}^n(\zeta)B_{p_1,m_1} \delta(2 p_1 + |m_1| - 2 p - |m|).
\end{align}

An important physical insight into the mechanism for Kerr beam self-cleaning can be gained by neglecting random linear mode coupling, and truncating the
coupled-mode equations~\eqref{Enumeric} to the first three transversal modes only, including the fundamental mode (i.e., we set $m=0$, $p=0,1,2$ in ~\eqref{Enumeric}. 
Let us define $A_p \equiv B_{p,0}$, where $A_i\equiv\sqrt{I_i}\exp(i\phi_i) (i=0,1,2)$. One obtains the conservation laws $I_0+I_1+I_2 = 1 $, and $I_0 - I_2 \equiv \alpha = const$, which provide the relations $I_{0,2}(\zeta) = (1-\eta\pm\alpha)/2$, where $I_1(\zeta)\equiv \eta < 1$.

\begin{figure}
\includegraphics[width=\linewidth]{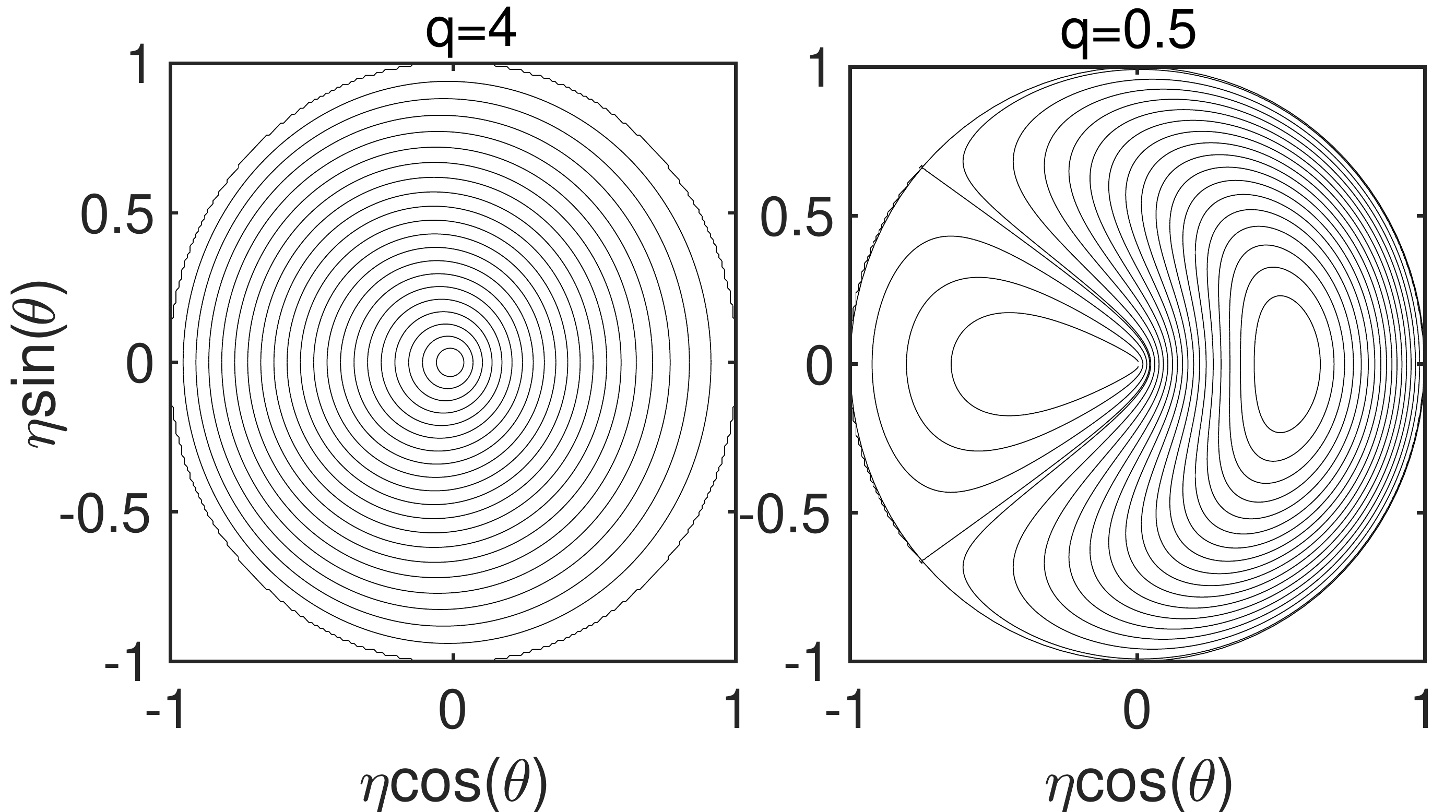}
\caption{Phase plane portraits for $q=4$ and $q=0.5$, with $\alpha=0$. \label{fig:contour}}
\end{figure}
From the truncated mode amplitude equations (see Eqs(1) of the Supplementary note), we obtain evolution equations for the normalized power $\eta$ of the intermediate transversal mode with $(m=0,p=1)$ and the phase $\theta=\phi_0+\phi_2-2\phi_1$
in compact Hamiltonian form 
\begin{equation}\label{H}
\frac{d\eta}{d\xi} = \frac{dH}{d\theta}, \frac{d\theta}{d\xi} = - \frac{dH}{d\eta},
\end{equation}
where $H=H_T\cos(\theta) +H_S$, $H_T= \eta\sqrt{(1-\eta)^2 - \alpha^2}, H_S= (q  - K_1)\eta  + K_2\eta^2$,
%\begin{gather*}
%H=H_T\cos(\theta) +H_S,\\ 
%H_T= \eta\sqrt{(1-\eta)^2 - \alpha^2}, H_S= (q  - K_1)\eta  + K_2\eta^2 
%\end{gather*}
%
and $q= 32 D\pi / p$, with constant coefficients $K_1$ and $K_2$ as defined in the Supplementary note. Eqs.\ref{H} have saddle points $\eta_e =1-|\alpha|$ and $\eta_e=0$ at large $q$.
In the linear limit (i.e., for $p \ll D$, or sufficiently large $q$), the linear wave vector mismatch suppresses the FWM-induced energy exchange between transversal modes. 

A bifurcation of the saddle points occurs when new maximum (or minimum) of $H(\theta, \eta)$ appears on the phase diagram.
Namely, when $dH/d\theta=0$, and $dH/d\eta =0$. For the detailed discussion of the bifurcation conditions of the saddle points, again we refer to the Supplementary note.
The consequence of the parametric instability in the dynamics of the spatial mode coupling process can be easily visualized with the help of phase plane portraits (i.e., constant level curves of the Hamiltonian $H$) of the solutions of Eqs.\ref{H}, as illustrated in Fig. \ref{fig:contour}. As can be seen, when diffraction dominates over FWM (e.g., when $q=4$), the modes virtually do not exchange energy but only experience a relative phase rotation. 
On the other hand, at high input power $p$, (e.g., with $q=0.5$), the intermediate mode with $(m=0,n=1)$ can get fully depleted. Moreover, a separatrix trajectory divides the phase plane in two regions with opposite sense of rotation of the phase $\theta$. The separatrix emanates from the saddle point (outer circle, points with $\eta=1$) and reaches values of $\eta\simeq0$ before asymptotically (i.e., for large distances) returning to the outer circle.
In real world quantities, the threshold for spatial beam cleaning is $p= 32 \pi D/q_{1\text{crit}}$, or
$P = 32\pi P_{sf} \lambda /2 n_0 q_{1\text{crit}} L_B \sim 10 \text{kW}.$

\begin{figure}
\includegraphics[width=0.8\linewidth]{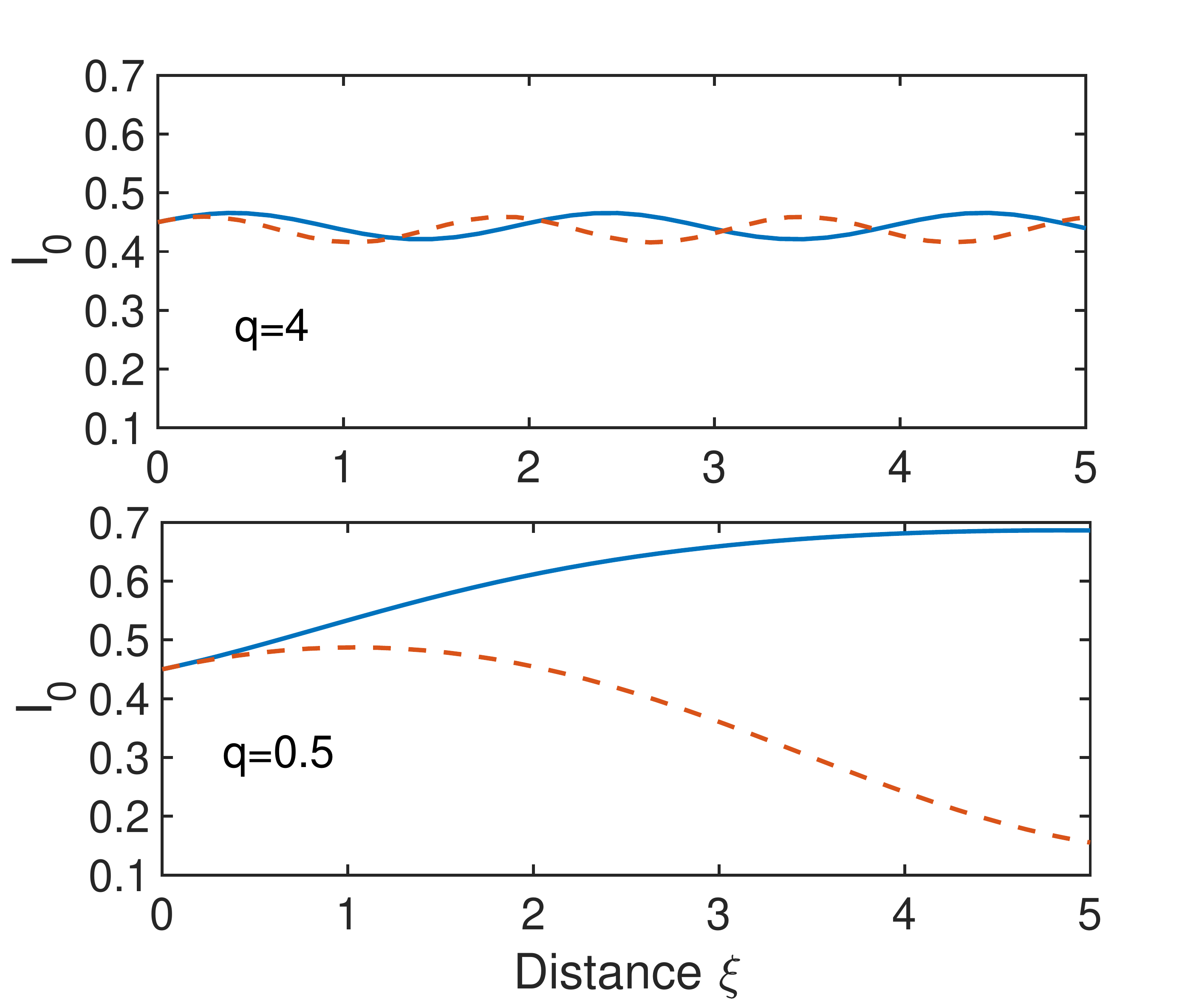}
\caption{Evolution of intensity fraction in fundamental mode with either $q=4$ or $q=0.5$, and $\alpha=0.4$, $\eta=0.5$ (blue solid curves) or $\alpha=0.1$, $\eta=0.2$ (dashed orange curves); $\theta(\xi=0)=0.35\pi$. \label{fig:power}}
\end{figure}

Parametric instability leads to a strong dependence of the spatial mode coupling process upon the input beam launching conditions into the MMF.  Fig. \ref{fig:power} shows that in linear conditions (i.e., with $q=4$) there is only a small variation in the intensity oscillation of the fundamental mode %(with $(m=0,n=0)$) 
as the input conditions are changed. Whereas in the presence of the parametric instability, there is a divergence of the flux of energy in and out of the fundamental mode as the input conditions are slightly modified. In Fig. \ref{fig:power}, we have considered two different input conditions, both corresponding to $45\%$ of initial intensity fraction in the fundamental mode. But in the case of the blue solid curves, the highest order mode is excited at the fiber input with only $10\%$ of the total input intensity. This corresponds to the case of an input beam with a relatively small transverse cross-section, so that most of the input intensity is coupled into the fundamental and the intermediate modes. Whereas the orange dashed curves in Fig. \ref{fig:power} correspond to a relatively larger fraction ($35\%$) of input intensity into the high-order transverse mode, as it occurs when the cross-section of the input beam gets larger.
The solid blue curve shows that the relative intensity in the fundamental mode grows up to $75\%$ upon propagation into the GRIN MMF. Whereas the orange curve shows that the fundamental mode is initially weakly amplified, but subsequently gets even depleted with distance. Thus we may conclude that self-cleaning should be accompanied by depletion of the intermediate mode, so that energy flows into both the fundamental and the high-order mode.

%In accordance with Eq~\eqref{E4}, in this case the intensity of $I_2$ grows larger with distance. At the same time, the fundamental intensity $I_0$ increases, and $I_1$  
%decreases with distance. 
%For example, if the amplitude of mode $|A_2|$ is small, then $|T|>|S|$, and $I_2$ starts to increase 
%along with the fundamental mode. 

This behavior can be extended to any triplets of modes, say $A_0,A_2, A_4$ or $A_1,A_2,A_3$. 
If the amplitude of the mode with largest wave number is small, its intensity will be increased during the 
propagation in GRIN fiber, alongside with the increase of amplitude of the mode with the smallest wave number.
This results in a transfer of energy to transversal modes with large wave numbers, together 
with the accumulation of energy in the mode with the smallest wave number (fundamental mode). 
When taking into account the conservation laws
\begin{align}\label{E7}
\sum\limits_{n=0}^\infty  f_n =1,\quad f_n &= \sum\limits_{p,m}|B_{p,m}|^2 \delta (n-2p-|m|), \nonumber \\   
%\end{equation}
%and
%\begin{equation}\label{E8}
\sum\limits_{n=0}^\infty n f_n &= \overline n =const,
\end{align}
one obtains that if the amplitude of modes with the largest wave number is increased, this means that the
amplitude of modes with intermediate wave numbers should decrease. At the same time, the fundamental mode intensity must also grow. In the process, 
the average mode number is conserved, as predicted by Eqs.\ref{E7}. This conservation also occurs in the presence of random mode coupling, as long as its spatial correlation length is much longer than the self-imaging period ($\simeq 0.6mm$, \cite{KrupaPRLGPI}) in the GRIN fiber.
The situation is fully analogous to that of 2D turbulence, which exhibits the two conservation laws: energy $E = \sum\limits_v \vec{V}_k^2$, and helicity $H = \sum\limits_v \vec{k}^2 \vec{V}_k^2$.
In hydrodynamic 2D turbulence, a direct cascade of helicity to large mode numbers (viscouse range) is accompanied by the inverse cascade of energy to the fundamental mode (condensate). 
At the end of the FWM cascade process, one should see the emergence of two spatial scales in the output beam: a condensate (very narrow fundamental mode) and a bottom layer (large number of HOMs with small amplitude), as it was predicted by a statistical study of Aschieri et al. \cite{PRA11b}.

\begin{figure}
	\includegraphics[width=0.49\linewidth]{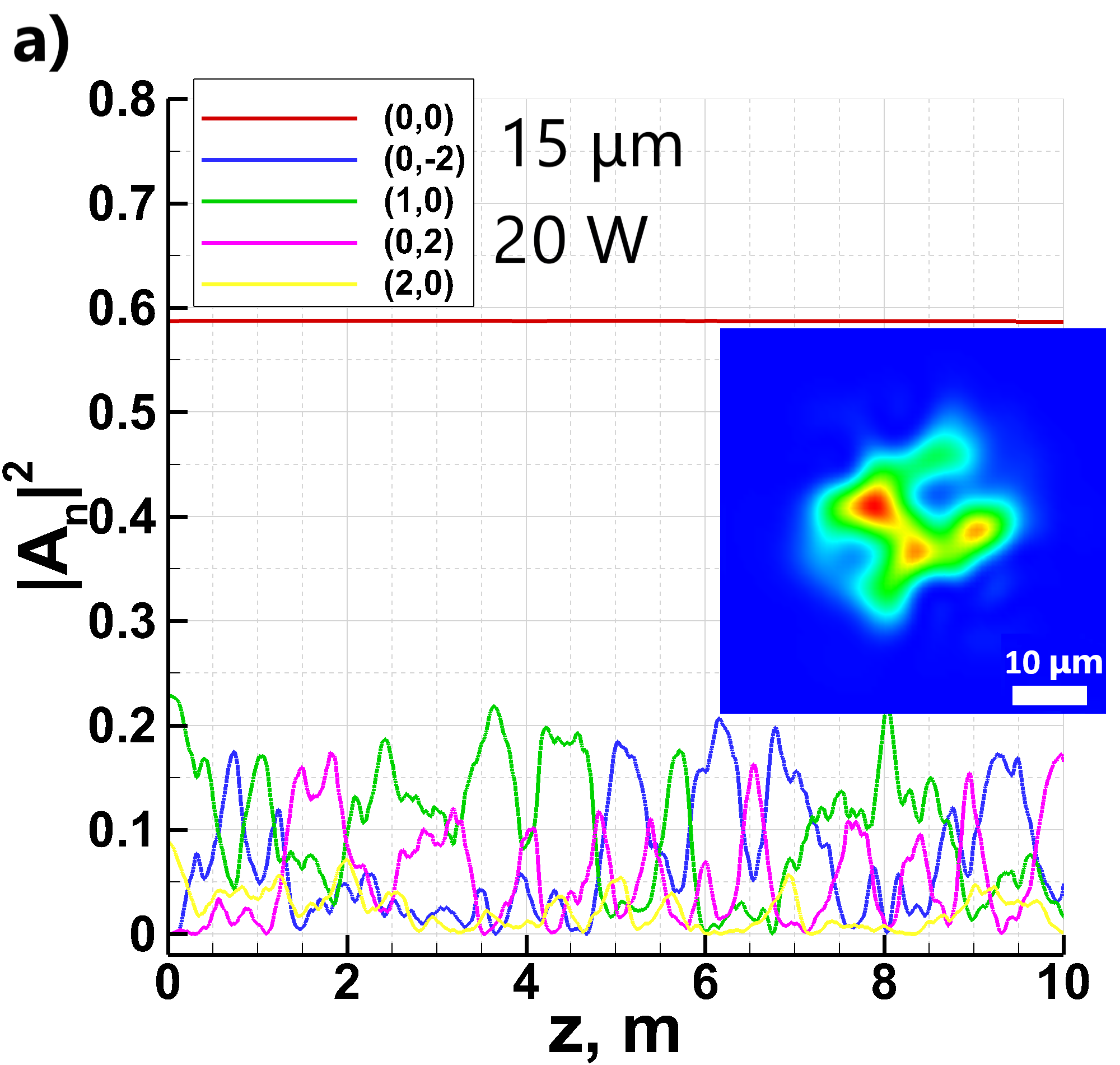}
	\includegraphics[width=0.49\linewidth]{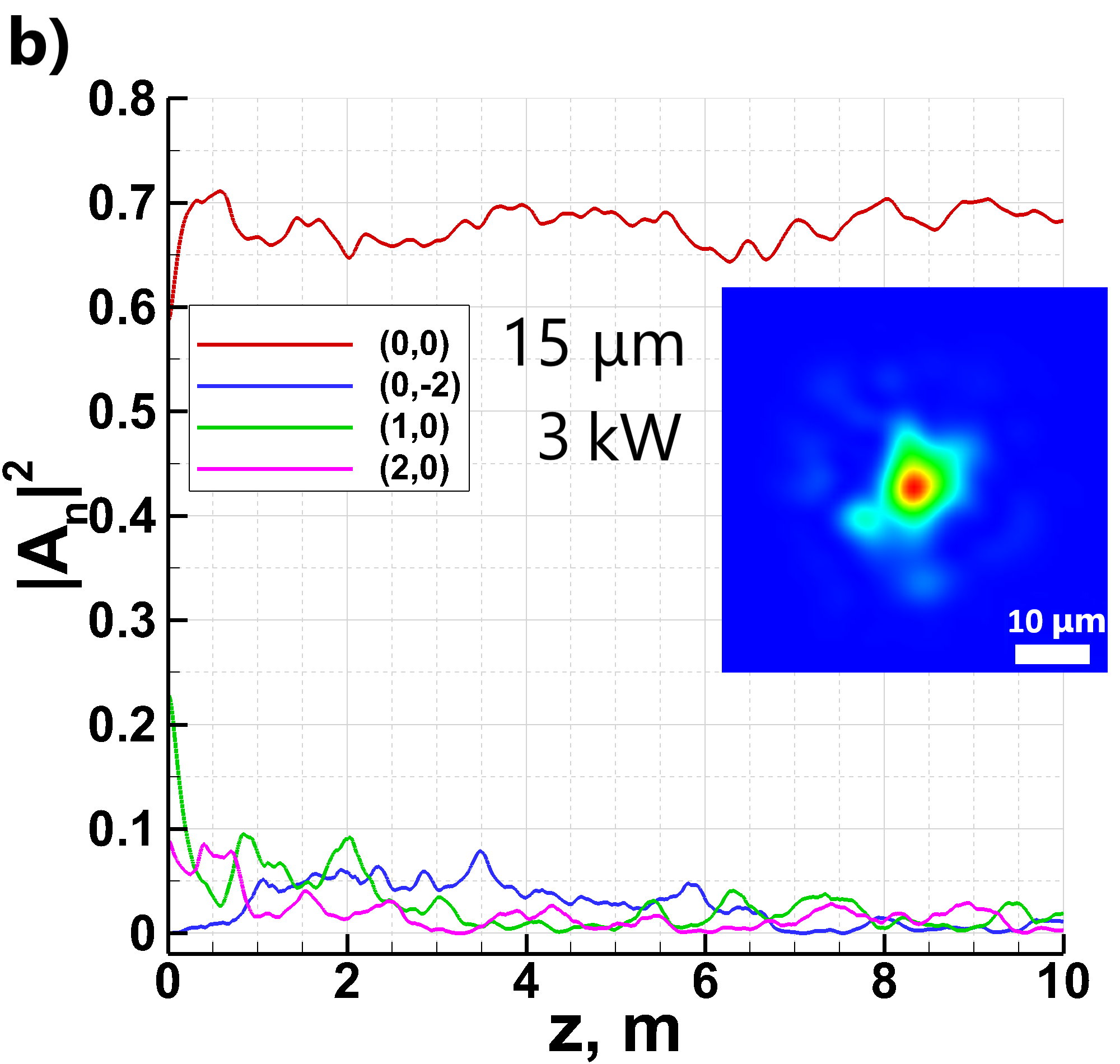}
	\par
	\includegraphics[width=0.49\linewidth]{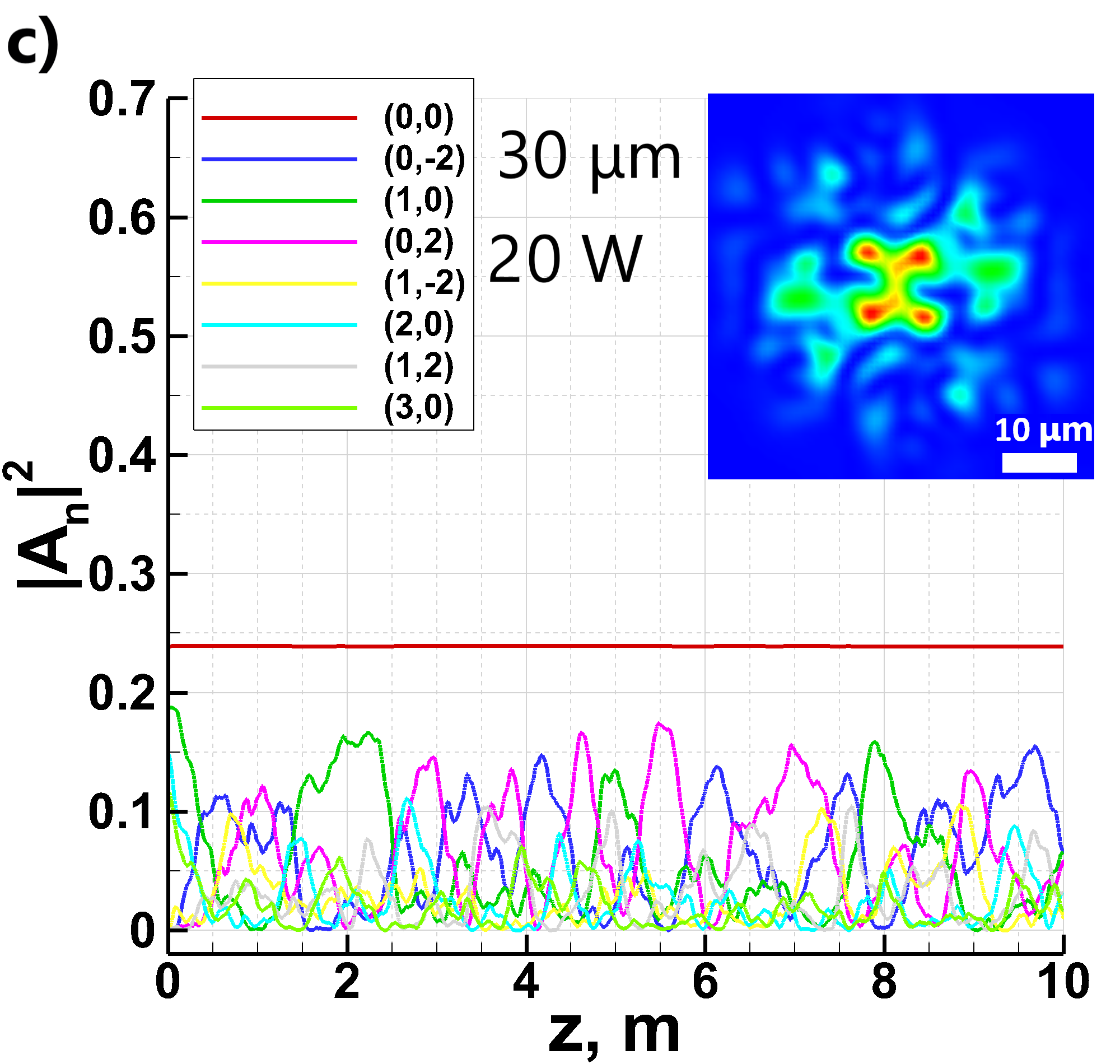}
	\includegraphics[width=0.49\linewidth]{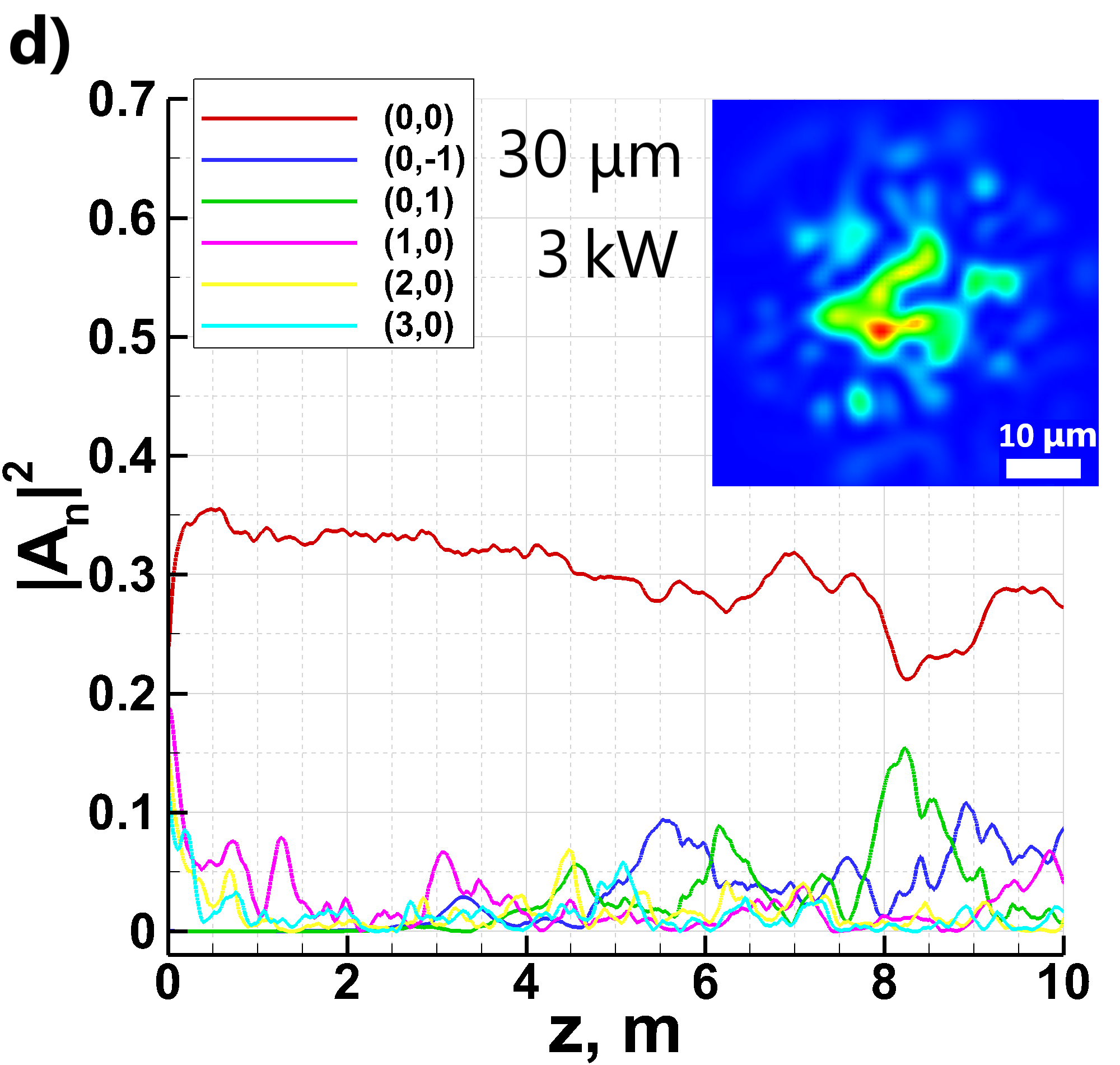}
	\caption{Evolution of mode intensities with fiber length at low ($P=20$~W, left) and high ($P=3$~kW, right) beam powers for 15~\mkm{} (a,b) and 30~\mkm{} (c,d) input beam radius. Insets~--- corresponding output intensity pattern.}
\label{fig:numerics1}
\end{figure}

To confirm our analytical treatment, we performed numerical simulation based on coupled-mode model~\eqref{Enumeric}.
The nonlinear coupling coefficients $f_{p,p_1,p_2,p_3}^{m,m_1,m_2,m_3}$ were determined by direct numerical calculation of the corresponding overlap integrals.
Coefficients $C_{p,p_1}^{m,m_1}$ correspond to random linear coupling between spatial modes with equal numbers $n=n1$.
In order to preserve the total power, it is necessary that the matrix $C=\{C_{p,p_1}^{m,m_1}\}$ be Hermitian.
We consider that only modes with the same mode numbers n are linearly coupled and that each element of the matrix $C$ is normally distributed with zero mean and varies randomly during propagating along the fiber with a correlation length of 10~cm.
As the initial data, a Gaussian beam of a certain radius was used, which was decomposed into modes, thereby setting the initial modal power distribution.
In our simulations, we only consider spatial modes with mode number $n\leq16$ (153 modes total).

\begin{figure}
\includegraphics[width=0.49\linewidth]{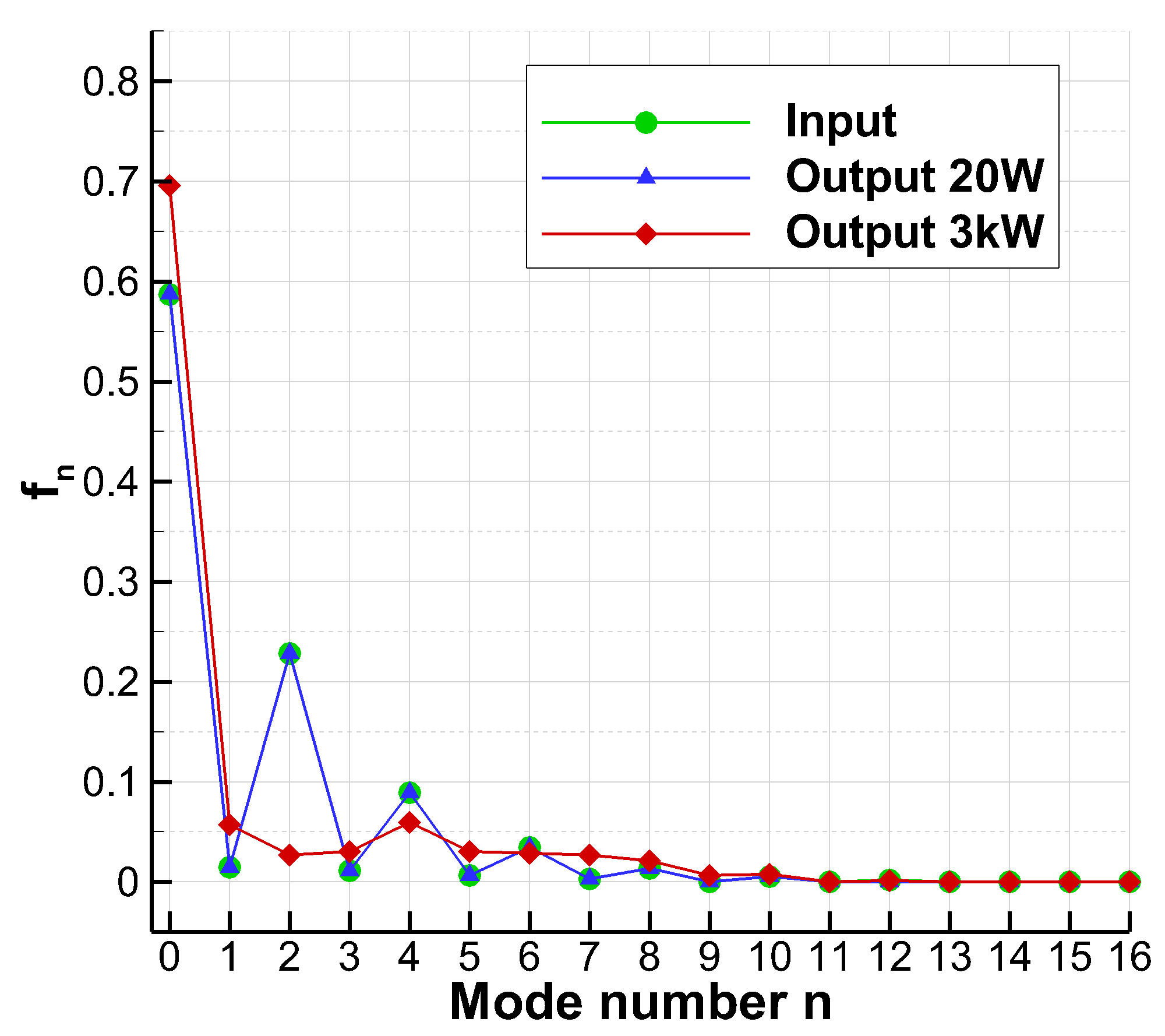}
\includegraphics[width=0.49\linewidth]{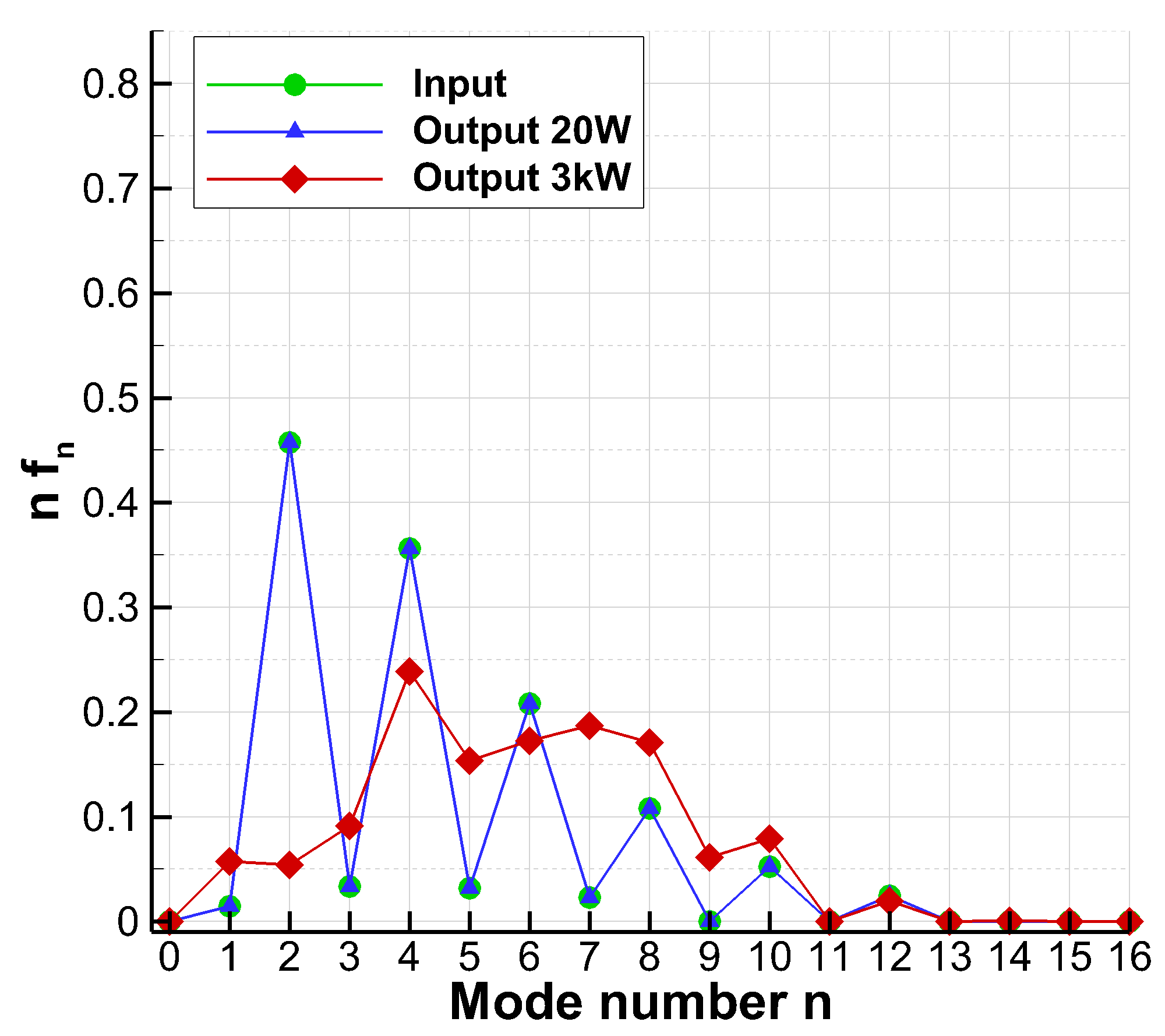}
	\caption{Evolution with fiber length of mode intensities $f_n$ and helicities $n f_n$ at low  ($P=10$~W) and high ($P=3$~kW) powers for 15~\mkm{} beam radius.}
\label{fig:numerics2}
\end{figure}

%This hypothesis has been confirmed by full numerical simulations, based on a coupled-mode expansion including all (radially symmetric and asymmetric) spatial modes.
%We included in the simulations random linear coupling among spatial modes with the same radial number $n$, and different azimuthal number $m$. 
In Figure~\ref{fig:numerics1}a we compare the evolution with distance along the fiber of the power in different fiber modes. In quasi-linear conditions (input beam power $20W$, Fig.~\ref{fig:numerics1}a for 15~\mkm{} beam radius), there is no power exchange between the fundamental mode (solid red curve) and HOMs.
However, random mode coupling occurs among all modes except for the fundamental, which leads to an irregular output intensity pattern (top right panel). Whereas for beam powers above the self-cleaning threshold (beam power $P=3$~kW, Figs.~\ref{fig:numerics1}b), parametric instability first increases the power fraction of the fundamental mode up to about $70\%$  (from $55\%$) within the first meter of fiber. Next, a continuous energy flow towards HOMs leads to nearly full depletion of intermediate LOMs along the fiber.
These properties are better clarified by plotting the corresponding spectra of modal intensities (energy) and of all mode number (helicity), respectively, as shown in Figure \ref{fig:numerics2}. As can be seen, above the threshold for spatial beam cleaning ($P=3$~kW), the LOMs get depleted, and energy flows towards the fundamental mode (inverse cascade) and the HOMs (direct cascade). 
For a larger beam (30~\mkm{}) the effect of self-cleaning becomes much weaker (Fig.~\ref{fig:numerics1}c,d).

\begin{figure}
	\includegraphics[width=0.75\linewidth]{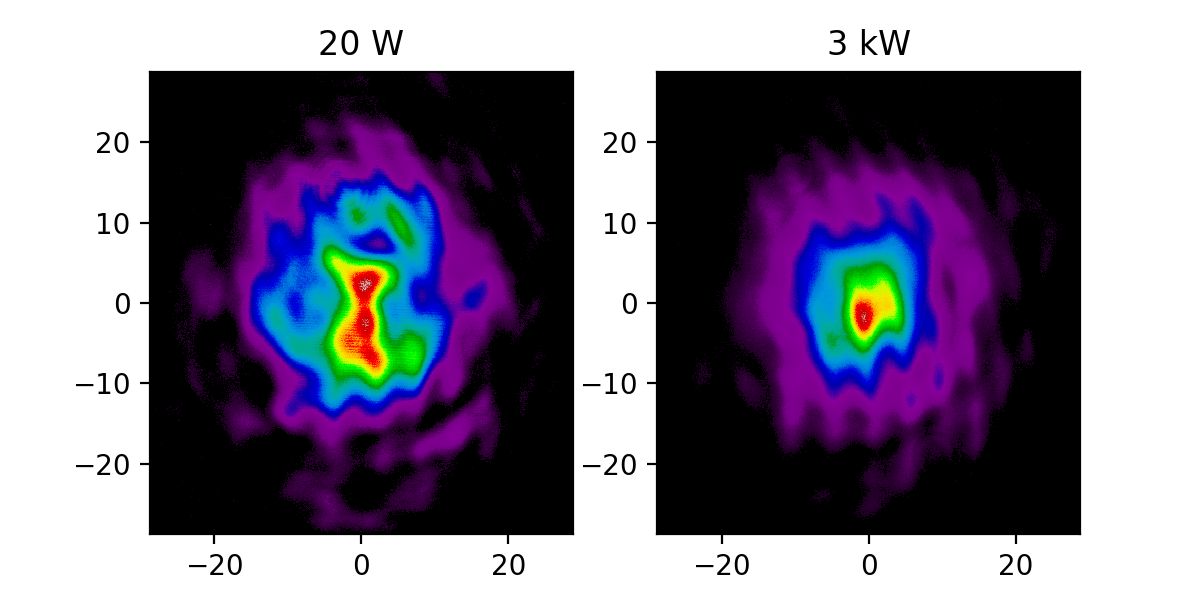}
	\includegraphics[width=0.75\linewidth]{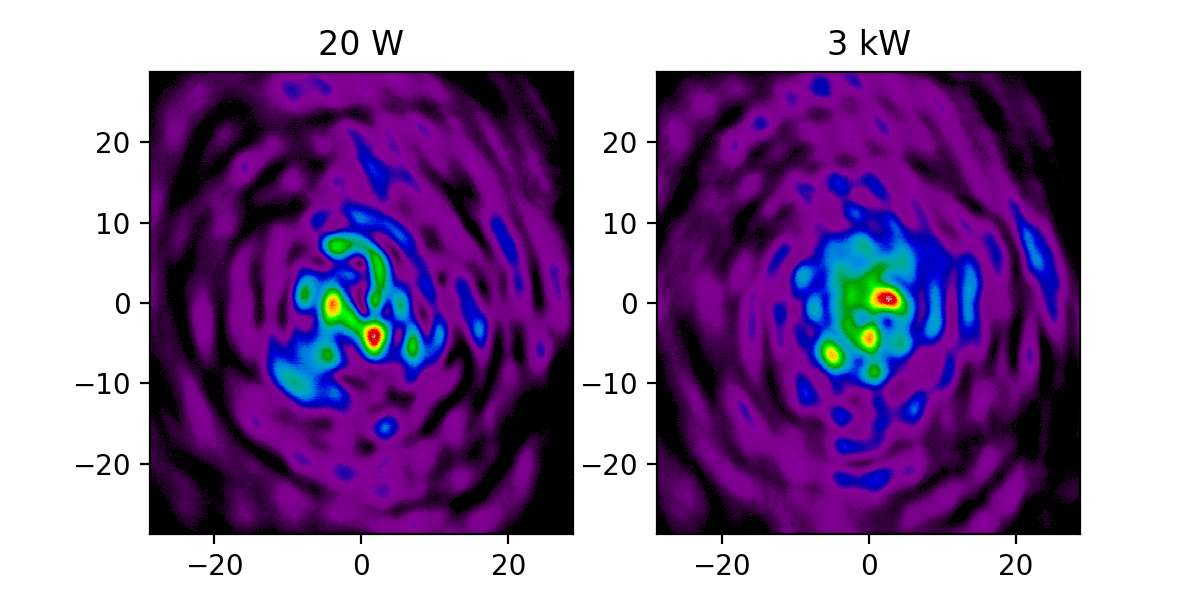}
	\caption{Experimental output intensity far field patterns for $17\pm1$~\mkm{} (top) and 30~\mkm{} (bottom) size input beams.}
\label{fig:exp1}
\end{figure}

In our experiments, we have verified the analytically and numerically predicted strong sensitivity, induced by the presence of a parametric instability, of beam self-cleaning upon variations of the input beam launching conditions (e.g., the input beam size, or the number of initially excited modes).
We launched into a 10-m-long MMF laser pulses from a Q-switched micro-chip Nd:YAG laser (Standa model STA-01-7), with a duration of 0.6~ns at the wavelength of 1064 nm.
Using a 62.5~\mkm{} core GRIN fiber, in combination with a beam expander (Standa model 10BE03 diffraction-limited Galilean type beam expander) that allowed us for a continuous change of the input beam radius from 15 to 32~\mkm{}, we found that both the establishment and the stability of the self-cleaning effect are strongly dependent of the initial beam parameters, as our theory predicts.

%Firstly, the self-cleaning effect quickly degrades for input beam angles with the longitudinal axis of the fiber that are equal or larger than 1 degrees. In such cases, we observed that the output spatial distribution changes as the peak power grows larger, however the central spot is not clean nor robust, as it jumps to different positions upon slightly touching or bending the fiber.
%An additional 
A critical parameter for the observation of stable self-cleaning is the beam size at the fiber input.
For an input transverse beam size which is much larger than the diameter of the fundamental mode of the GRIN MMF leads to huge coupling losses, as too many HOMs are excited. When the input beam size is small enough, one excites a beam close to the fundamental mode of the fiber (of $\sim$10~\mkm{} radius).
Under such conditions, the output beam maintains a good quality, almost independently of power.
The most interesting case corresponds to an intermediate beam size, which is close to the condition for optimal coupling into the MMF, and still it remains substantially larger than the fundamental mode.
In Fig. \ref{fig:exp1} we illustrate typical cases for the power dependence of the output beam intensity pattern from the GRIN MMF.
As can be seen, by increasing the input beam radius from $17$~\mkm{} up to $30$~\mkm{} (correspondingly, the beam coupling efficiency drops from 90\% to 55\%)  while keeping the output peak power stable at $3$~kW, the beam self-cleaning effect is totally suppressed,
in accordance with theoretical predictions of Fig.~\ref{fig:numerics1}(c,d).
Whereas, for an input beam radius between 15 and 17($\pm1$)~\mkm{}, the speckled structure observed at low powers transforms, at powers ${>}1$~kW, into a bright spot (Fig.~\ref{fig:exp1}) that is totally insensitive to fiber deformations, and changes of the input beam polarization state \cite{Krupanatphotonics}. Here the input beam radius is measured at the intensity level of $1/e^2$ from the peak.

\begin{figure}
\includegraphics[width=\linewidth]{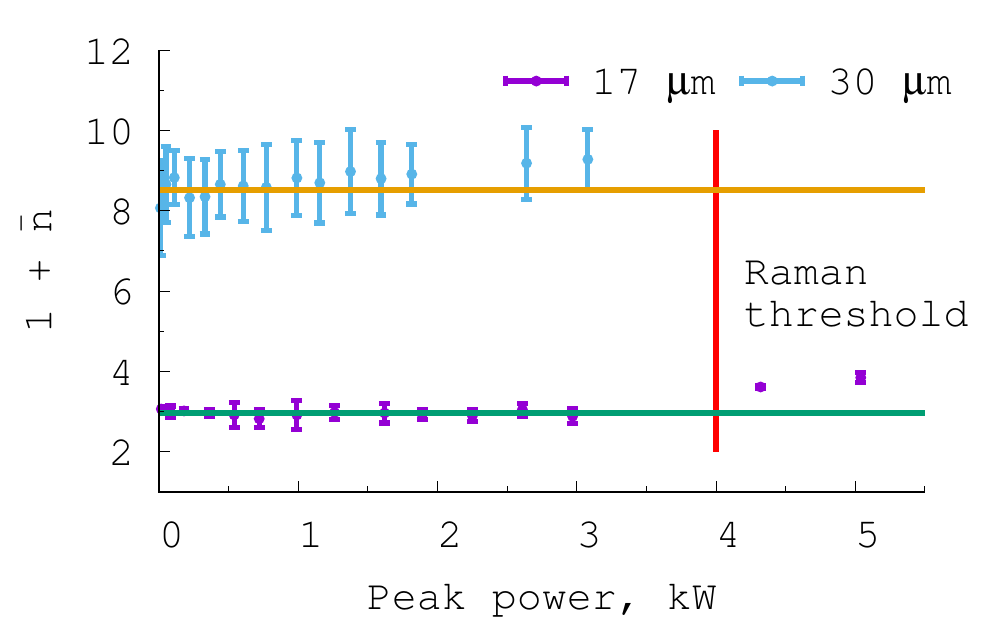}
	\caption{Experimental (dots) and theoretical (solid curves) dependence of the average mode number (from Eq.~\eqref{eq:navg}) vs. output peak power, for $17\pm1$~\mkm{} and 30~\mkm{} size input beams.}
\label{fig:exp2}
\end{figure}

Our main result is the experimental confirmation of the conservation of the average mode number, as predicted by Eqs.\ref{E7}, in the process of beam self-cleaning. This condition characterizes the simultaneous occurrence of inverse and direct energy cascade in hydrodynamic 2D turbulence \cite{kraichnan,kandm}. 
In the experiment, we measure the average mode number as
\begin{equation}
	\bar{n} + 1 =
	\frac{1}{2} \left( \frac{S_{\text{NF}}}{S^0_{\text{NF}}} + \frac{S_{\text{FF}}}{S^0_{\text{FF}}} \right) =
	\frac{\int{ r^2 |A(r)|^2 d\vec{r}} + \int{ |\vec{\nabla} A|^2 d\vec{r}}}{2\int{ |A(r)|^2 d\vec{r}}}
\label{eq:navg}
\end{equation}
where: $S^0$ is the area of the fundamental mode %(mean $r^2$)
, whereas $S_\text{NF}$, $S_\text{FF}$ denote the beam area in near field and far field, respectively.
The $S^0$ value in the near- and far-field was calculated by using the ABCD matrix method, in order to normalize integrals in the equation for the average
mode number. The theoretically calculated size of the fundamental mode for a 62.5~\mkm{} core GRIN fiber was used as initial value. 
As a result, we were able to calculate the magnification factor in both cases.
In Fig. \ref{fig:exp2} we show the measured dependence of the average mode number for the two cases of $17\pm1$~\mkm{} (top) and 30~\mkm{} (bottom) size input beams. As can be seen, the input average mode number is conserved in all cases, in excellent agreement with the theoretical values (solid curves). Note that the total number of modes in the GRIN MMF is equal to $M=(n+1)(n+2)/2$.
%at least until the power threshold for stimulated Raman scattering. 
Therefore, our experiments confirm that the mechanism of self-cleaning is based on a mode redistribution from intermediate transversal modes to the fundamental (inverse cascade), and to higher-order ones (direct cascade), as in hydrodynamic 2D-turbulence.
Such a behavior is associated with the presence of a parametric spatial instability in the nonlinear mode coupling process.

To summarize, we experimentally demonstrated that beam self-cleaning in multimode optical fibers is a process that conserves the average mode number, in spite of the dramatic nonlinear change and self-organization into the fundamental mode of the output intensity pattern. We pointed out that the process of energy flow into the fundamental and HOMs at the expense of modes with intermediate wave numbers originates from a parametric instability, in analogy with hydrodynamic 2D turbulence. These results provide yet another demonstration of the parallelisms between hydrodynamic and optical turbulence, and of the universality of mechanisms for spatial pattern generation in different physical settings.
 
\begin{acknowledgments}
This work was supported by the Russian Ministry of Science and Education (Grant 14.Y26.31.0017), and the European Research Council (ERC) under the European
Union's Horizon 2020 research and innovation programme (grant No. 740355). Work of SKT was supported by the Russian Science Foundation (Grant No. 17-72-30006). \end{acknowledgments}
 
\bibliography{BIBLIO_MultimodeTemp3}

%\begin{thebibliography}{99}
%\bibitem{1d} z
%\end{thebibliography}

\end{document}